\documentclass[aps,prd,twocolumn,showpacs,floatfix,preprintnumbers%
,nofootinbib,superscriptaddress]{revtex4-1}

\usepackage{graphicx}
\usepackage{color}

\newcommand{\exclude}[1]{}

\begin{document}

\title{The meV mass frontier of axion physics}

\author{Georg~G.~Raffelt}
\affiliation{Max-Planck-Institut f\"ur Physik
  (Werner-Heisenberg-Institut),
  F\"ohringer Ring~6, 80805 M\"unchen, Germany}

\author{Javier Redondo}
\affiliation{Max-Planck-Institut f\"ur Physik
  (Werner-Heisenberg-Institut),
  F\"ohringer Ring~6, 80805 M\"unchen, Germany}

\author{Nicol\'as Viaux}
\affiliation{Departamento de Astronom{\'\i}a y Astrof{\'\i}sica,
Pontificia Universidad Cat{\'o}lica de Chile, Av.\ Vicu{\~n}a
Mackenna 4860, 782-0436 Macul, Santiago, Chile.}

\date{19 August 2011}
\preprint{MPP-2011-11}

\begin{abstract}
We explore consequences of the idea that
the cooling speed of white dwarfs 
can be interpreted in terms of axion emission. In this case 
the Yukawa coupling
to electrons has to be
$g_{ae}\sim1\times10^{-13}$, corresponding to an axion mass
of a few~meV. Axions then provide only a small fraction of the
cosmic cold dark matter, whereas core-collapse supernovae release a
large fraction of their energy in the form of axions. We estimate
the diffuse supernova axion background (DSAB) in the universe,
consisting of 30~MeV-range axions with a radiation density
comparable to the extra-galactic background light. The DSAB would be
challenging to detect. However, axions with white-dwarf
inspired parameters can be accessible in a next
generation axion helioscope.
\end{abstract}


\maketitle

\section{Introduction}

The Peccei-Quinn (PQ) mechanism remains
perhaps the most compelling explanation for the absence of
CP-violating effects from the QCD vacuum
structure~\cite{Peccei:2006as, Kim:2008hd, Nakamura:2010zi}. An
unavoidable consequence is the existence of the axion, the
Nambu-Goldstone boson of a new U(1)$_{\rm PQ}$ symmetry. Axions
acquire a mass $m_a \sim m_\pi f_\pi/f_a$ by their mixing with
neutral mesons, where $m_\pi=135$~MeV and $f_\pi=92$~MeV are the
pion mass and decay constant, and $f_a$ is a large energy scale
related to the spontaneous breaking of U(1)$_{\rm PQ}$. Axions
generically interact with hadrons and photons. They may also
interact with charged leptons, the DFSZ model~\cite{Dine:1981rt}
being a generic case. All interactions are suppressed by $f_a^{-1}$,
so for large $f_a$, axions are both very light and very weakly
interacting. Reactor and beam-dump experiments require $m_a\alt
30$~keV \cite{Nakamura:2010zi}, while precision cosmology excludes
the $m_a$ range 1~eV--300~keV \cite{Hannestad:2010yi}. Sub-eV mass
axions would still be copiously produced in stars. The cooling of
white dwarfs (WDs), neutron stars and supernova (SN) 1987A pushes
the limits to $m_a\alt10$~meV \cite{Raffelt:2006cw}, i.e.\
$f_a\agt10^9$~GeV. We here explore the impact of axions near this
limit, i.e.\ the meV frontier of axion physics.

This range is complementary to the other extreme of the allowed
axion window. During the QCD epoch of the early universe, the axion
field gets coherently excited, generating a cold dark matter (CDM)
fraction of $\rho_a/\rho_{\rm CDM}\sim\Theta_{\rm i}^2\,(10~\mu{\rm
eV}/m_a)^{1.2}$ \cite{Sikivie:2006ni}, where \hbox{$\Theta_{\rm
i}=a_{\rm i}/f_a$} is the initial ``misalignment angle'' relative to
the CP-conserving value. For $\Theta_{\rm i}\sim 1$, axions with
$m_a\sim10~\mu$eV ($f_a\sim 10^{12}$~GeV) provide all of CDM and can
be detected in the ADMX experiment~\cite{Asztalos:2009yp}. If the
reheating temperature after inflation was large enough to restore
the PQ symmetry, our visible universe emerges from many domains and
an average $\langle \Theta_{\rm i}^2\rangle \sim\pi^2/3$ has to be
used. In this case, axions also emerge from the decay of topological
defects and the CDM density could correspond to $m_a$ as large as a
few 100~$\mu$eV~\cite{Wantz:2009it}. Either way, meV-mass axions
provide only a subdominant CDM component.

\section{Cooling of compact stars}

The most restrictive astrophysical
limits on those axion models that couple to charged leptons arises
from WDs. An early study used the WD cooling speed, as manifested in
their luminosity function, to derive a limit on the axion-electron
coupling of $g_{ae}\alt4\times10^{-13}$ \cite{Raffelt:1985nj}. In
the early 1990s it became possible to test the cooling speed of
pulsating WDs, the class of ZZ Ceti stars, by their measured period
decrease $\dot P/P$. In particular, the star G117-B15A was cooling
too fast, an effect that could be attributed to axion losses if
$g_{ae}\sim2\times10^{-13}$ \cite{Isern:1992}. Over the past twenty
years, observations and theory have improved and the G117-B15A
cooling speed still favors a new energy-loss
channel~\cite{Isern:2010wz}. What is more, the WD luminosity
function also fits better with axion cooling if
$g_{ae}=0.6$--$1.7\times10^{-13}$ \cite{Isern:2008nt}.

While complete confidence in this intriguing interpretation is
certainly premature (perhaps even in the need for a novel WD cooling
itself), the required axion parameters are very specific, motivating
us to explore other consequences based on the WD benchmark.

Axion cooling of SNe has been widely discussed in the context of
SN~1987A \cite{Raffelt:2006cw, Raffelt:1987yt, Janka:1995ir,
Hanhart:2000ae}. The 10~s duration of the neutrino burst supports
the current picture of core collapse and cooling by quasi-thermal
neutrino emission from the neutrino sphere. New particles that are
more weakly interacting than neutrinos, such as the axions discussed
here, can be produced in the inner SN core, leave unimpeded, and in
this way drain energy more efficiently than neutrinos, which can
escape only by diffusion. The SN~1987A neutrino burst duration
precludes a dominant role for axions. Quantitatively, this argument
depends on the model-dependent axion-nucleon couplings, the
uncertain emission rate from a dense nuclear medium, and on sparse
data. As we shall see, the limit does not preclude the WD
interpretation, but a SN would lose a significant fraction of its
energy in the form of axions.

The speed of neutron-star cooling as measured by the surface
temperature of several pulsars~\cite{Page:2005fq} is another
possible laboratory to search for axion cooling and a limit
comparable to the SN~1987A bound was found~\cite{Umeda:1997da}.
However, neutron-star cooling depends even more dramatically on
nuclear physics uncertainties and on the details of axion-nucleon
coupling than the SN~1987A bound so that it is hard to make such
arguments precise. However, if the WD interpretation applies, axion
emission is another effect to be taken into account in the
complicated theory of neutron-star cooling.

\section{Diffuse SN axion background}

Returning to the energy
loss of proto neutron stars after core collapse, axions saturating
the SN~1987A limit are emitted as copiously as neutrinos.
Then one not only expects a strong axion burst
from each SN, but also a large cosmic diffuse background flux from
all past SNe, the diffuse SN axion background (DSAB) in analogy to
the diffuse SN neutrino background (DSNB) \cite{Beacom:2010kk}. All
past SNe in the universe provide a local $\bar\nu_e$ flux of order
$10~{\rm cm}^{-2}~{\rm s}^{-1}$ \cite{Beacom:2010kk} that will become
detectable in a Gd-enriched version of
Super-Kamiokande~\cite{Watanabe:2008ru} or a future large
scintillator detector~\cite{Autiero:2007zj} with a rate of a few
events per year. The estimated core-collapse rate is scaled to the
amount of extra-galactic background light (EBL), representing the
integrated star-formation history~\cite{Horiuchi:2008jz}. The
intensity of the EBL is 50--$100~{\rm nW}~{\rm m}^{-2}~{\rm
ster}^{-1}$, corresponding to an energy density of 13--$26~{\rm
meV}~{\rm cm}^{-3}$, i.e.\ about 10\% of the energy density provided
by the cosmic microwave background.

The present-day average core-collapse rate is $R_{\rm
cc}=1.25\times10^{-4}~{\rm Mpc}^{-3}~{\rm yr}^{-1}$ and increases
with redshift roughly proportional to $10^z$ until $z=1$ and then
flattens or slightly decreases \cite{Horiuchi:2008jz}. Assuming that
every SN releases $3\times10^{53}$~erg in the form of neutrinos of
all flavors and integrating over $R_{\rm cc}$, properly redshifting
the energy, leads to a present-day DSNB of $26~{\rm meV}~{\rm
cm}^{-3}$, almost identical with the EBL. In other words, stellar
populations release on average as much gravitational binding energy
in the form of neutrinos as they release nuclear binding energy in
the form of photons.

For meV-mass axions, therefore, the energy density of the DSAB can be
comparable to the DSNB and the EBL, and indeed would be the most
important axion population in the universe. The axion losses of
ordinary stars would contribute a much smaller energy density, just
as the neutrinos emitted by all ordinary stars contribute an energy
density of only about 7\% of the EBL~\cite{Porciani:2003zq,
Fukugita:2004ee}.

\begin{figure}
\includegraphics[width=0.70\columnwidth]{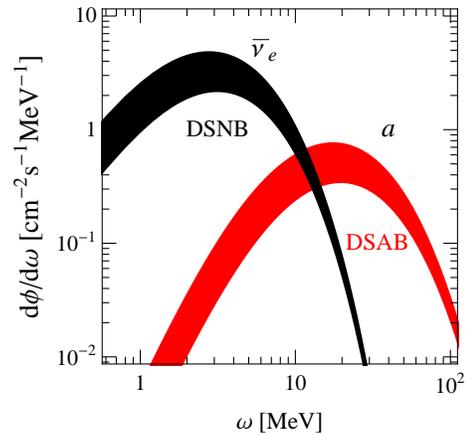}
\vspace{-.3cm}
\caption{\small Diffuse backgrounds of
SN neutrinos and axions, assuming
that either one carries away the full SN energy ($3\times10^{53}~{\rm
erg}$). The width of the bands reflects only the
uncertainty in the core-collapse rate $R_{\rm cc}$. For $\bar\nu_e$
a thermal spectrum with $T=4$~MeV is assumed, carrying away 1/6 of
the total energy, whereas for axions we use the
bremsstrahlung-inspired spectrum of Eq.~(\ref{eq:spectrum}) with
$T_{\rm core}=30$~MeV.\label{fig:DSNB}}
\end{figure}

The DSAB will be calculated in analogy to the DSNB where the
$\bar\nu_e$ spectrum as a function of present-day $\bar\nu_e$ energy
$E$ is given by the redshift integral
\begin{equation}
\frac{dN}{dE} =
\int_0^\infty dz\,
\Bigl\lbrace(1+z)\,\varphi[E (1 + z)] \Bigr\rbrace
\Bigl\lbrace R_{\rm cc}(z)\left|\frac{dt}{dz}\right|\Bigr\rbrace\,,
\end{equation}
where $R_{\rm cc}$ is the core-collapse rate. The function
$\varphi(E')$ provides the number of $\bar\nu_e$ per rest-frame
energy interval $dE'$ released by an average SN. Notice that $N_{\rm
tot}=\int dE\, (1+z)\varphi[E(1+z)]$, the total number of neutrinos
released by a SN, is invariant against redshift. Further,
$|dt/dz|^{-1} = H_0 (1+z) [\Omega_\Lambda + \Omega_{\rm
M}(1+z)^3]^{1/2}$ with cosmological parameters taken as $H_0 =
70~{\rm km}~{\rm s}^{-1}~{\rm Mpc}^{-1}$, $\Omega_\Lambda = 0.7$,
and $\Omega_{\rm M} = 0.3$. Notice that $|dt/dz|$ and $R_{\rm cc}$
actually form one combined factor proportional to the ratio of the
average luminosity per galaxy in SN neutrinos relative to stellar
photons. Assuming $E_{\rm tot}=3\times10^{53}~{\rm erg}$, 1/6 of
this in $\bar\nu_e$, and $T_{\bar\nu_e}=4$~MeV after flavor
oscillations~\cite{Fischer:2009af}, we show the DSNB in
Fig.~\ref{fig:DSNB}. The width of the band reflects only the
uncertainty of $R_{\rm cc}$, not the uncertainty of SN neutrino
emission.

\section{Axion properties}

Before estimating the DSAB we
summarize the relevant phenomenological axion properties. Their mass
is
\begin{equation}\label{eq:axionmass}
m_a=\frac{\sqrt{z}}{1+z}\,\frac{m_\pi f_\pi}{f_a}
=6~{\rm meV}\,\frac{10^9~{\rm GeV}}{f_a}\,,
\end{equation}
where $z=m_u/m_d=0.35$--0.60~\cite{Nakamura:2010zi}, but we always
use the canonical value $z=0.56$. The interaction with fermion $f$
has the axial-vector derivative structure
\begin{equation}
{\cal L}_{af}=(C_f/2f_a)\,\bar\Psi_f\gamma^\mu\gamma_5\Psi_f\,\partial_\mu
a\,,
\end{equation}
where $C_f$ is a numerical coefficient and $g_{af}=C_f m_f/f_a$ the
corresponding Yukawa coupling. For protons $C_p=[C_u-1/(1+z)]\Delta
u+ [C_d-z/(1+z)]\Delta d$ and neutrons $C_n=[C_u-1/(1+z)]\Delta
d+[C_d-z/(1+z)]\Delta u$, where $\Delta u=0.84\pm0.02$ and $\Delta
d=-0.43\pm0.02$ \cite{Nakamura:2010zi}. The axion-photon interaction
is
\begin{equation}
{\cal L}_{a\gamma}=
-(g_{a\gamma}/4)F_{\mu\nu}\tilde
F^{\mu\nu}\,a=g_{a\gamma}{\bf E}\cdot{\bf B}\,a
\end{equation}
with $g_{a\gamma}=\alpha/(2\pi f_a)\,[E/N-2/3(4+z)/(1+z)] \approx
\alpha/(2\pi f_a)\,(E/N-2)$ and $E/N$ is the ratio of the
electromagnetic and color anomalies.

We assume DFSZ axions~\cite{Dine:1981rt} for which $E/N=8/3$,
$C_u=\frac{1}{3}\,\sin^2\beta$ and
$C_d=C_e=\frac{1}{3}\,\cos^2\beta$. The WD value $g_{ae}= 1\times
10^{-13}$ for the axion-electron coupling implies
\begin{equation}
f_a=1.7\ \cos^2\beta\ 10^9~{\rm GeV}\,.
\end{equation}
We choose $\cos^2\beta=1/2$ because much smaller
values would favor SN emission over WD emission and lead to
overly optimistic DSAB estimations. This choice implies
$f_a=0.85\times10^9~{\rm GeV}$, $m_a=7$~meV, $C_e=1/6$, $C_n=0$
and $C_p=-1/3$. Then
$g_{a\gamma}=1.0\times10^{-12}~{\rm GeV}^{-1}$ is two orders
below the CAST sensitivity~\cite{Andriamonje:2007ew}.
The axion-nucleon couplings are $g_{an}=0$ and
$g_{ap}=3.7\times10^{-10}$.

\section{Estimating the DSAB}

For such small couplings, axions escape
freely from a SN core once produced. The dominant production process
is nucleon bremsstrahlung $NN\to NNa$, but a reliable calculation
has proven elusive~\cite{Raffelt:2006cw}. Axions couple to the
nucleon spin and are produced in spin fluctuations caused by the
tensor force in $NN$ collisions. Early calculations used
nondegenerate free nucleons and a one-pion exchange (OPE) potential
in Born approximation. However, if the nucleon spin-fluctuation rate
$\Gamma_\sigma$ were as large as found in these calculations,
destructive interference from multiple scattering would reduce the
emission rate~\cite{Janka:1995ir}. On the other hand, based on
measured $NN$ scattering data, the spin-flip cross section was found
to be much smaller than implied by the OPE
approximation~\cite{Hanhart:2000ae}.

For a phenomenological description of the axion interaction with a
nuclear medium we use the dynamical spin-density structure function
$S_\sigma(\omega)$~\cite{Janka:1995ir, Hanhart:2000ae}. The
absorption rate for axions of energy $E_a$ is
$g_{aN}^2\,(\rho/8m_N^3)\,\omega\,S_\sigma(E_a)$ where $\rho$ is the
matter density.
If we
ignore spin correlations between different nucleons, the
normalization is $\int_{-\infty}^{+\infty}S_\sigma(\omega)\,d\omega/(2\pi)=1$.
Emission and absorption are related by detailed balancing, implying
$S_\sigma(-\omega)=S_\sigma(\omega)\,e^{-\omega/T}$.
The spectral axion emission per unit volume is therefore
\begin{equation}
\frac{dQ}{d E_a} = \frac{g_{aN}^2 \rho}{16\,\pi^2}\,
\frac{E_a^4}{m_N^3}\,S_\sigma(-E_a)\,.
\end{equation}
The energy-loss rate per unit mass $Q/\rho$ is given by $\epsilon_a
= (g_{aN}^2/8\pi)\,(T^4/m_N^3)\,F$ in terms of the dimensionless
integral $F=\int_0^\infty
S_\sigma(-\omega)\,(\omega/T)^4\,d\omega/2\pi$.

Low-energy bremsstrahlung is essentially a classical phenomenon. Classical
spins kicked by a random force with a rate $\Gamma_\sigma$ imply
$S_\sigma =\Gamma_\sigma/(\omega^2+\Gamma_\sigma^2/4)$ and inspire a
one-parameter representation fulfilling all requirements
\begin{equation}
S_\sigma(\omega)=\frac{\Gamma_\sigma}{\omega^2+\Gamma_\sigma^2/4}\,
\frac{2}{e^{-\omega/T}+1}\,.
\end{equation}
We consider only interactions with protons (typical abundance 30\%
per baryon) and adopt $F=1$ as a rough estimate so that
$\epsilon_a\sim g_{ap}^2\,1.6\times10^{37}~{\rm erg}~{\rm
g}^{-1}~{\rm s}^{-1}(T/30~{\rm MeV})^4$. The SN~1987A neutrino
signal duration requires $\epsilon_a\alt 1\times10^{19}~{\rm
erg}~{\rm g}^{-1}~{\rm s}^{-1}$, providing
$g_{ap}\alt0.8\times10^{-9}$ \cite{Raffelt:2006cw}.

Our assumptions correspond to $\Gamma_\sigma\sim 2T$. Assuming an
isothermal SN core, the axion spectrum is
\begin{equation}\label{eq:spectrum}
\frac{dn}{d E_a}\propto \frac{E_a^3}{E_a^2+T^2}\,
\frac{2}{e^{E_a/T}+1}\,.
\end{equation}
This distribution is of course not very well determined and mostly
serves the purpose of illustration. With $T_{\rm core}=30$~MeV we
find $\langle E_a\rangle\sim 80$~MeV.

For our WD inspired axion parameters, a SN emits roughly 1/8 of its
energy as axions. Considering all uncertainties, this fraction could
be smaller or as large as 1/2, at which point it would seriously
affect the SN~1987A signal. Assuming all energy is emitted in axions
and with the spectral shape of Eq.~(\ref{eq:spectrum}) we find the
DSAB shown in Fig.~\ref{fig:DSNB}. The average axion energy is about
35~MeV.

\section{Detecting the DSAB?}

Detecting this flux is extremely
challenging. Axions with the parameters considered here interact
much more weakly than neutrinos of comparable energy. The DSNB will
be detectable in Super-Kamiokande and in next generation large-scale
detectors, but the DSAB produces a much smaller signal.

One may think that conversion in large-scale astrophysical magnetic
fields may provide a detectable signal. It would have to stick above
the diffuse gamma-ray background in the 30~MeV region that was
measured by the EGRET satellite to be 1--$2\times10^{-6}\ {\rm
cm}^{-2}\ {\rm s}^{-1}~{\rm sr}^{-1}~{\rm MeV}^{-1}$
\cite{Strong:2004ry}. Dividing the DSAB in Fig.~\ref{fig:DSNB} by
$4\pi$ to obtain a flux per sterad, we see that the conversion
probability would have to be of order $10^{-4}$.

In a transverse $B$ field and after travelling a distance $L$, the
axion-photon oscillation probability is
\begin{equation}
P_{a\to\gamma}=(g_{a\gamma} B/q)^2\,\sin^2(q L/2)\,,
\end{equation}
where $q=(m_a^2-m_\gamma^2)/2E$. For $m_a=7$~meV we can neglect the
photon plasma mass in interstellar space. For $E=30$~MeV the
oscillation length $4\pi E/m_a^2$ is 1500~km. For these parameters,
the maximum conversion rate is
$P_{a\to\gamma}=6\times10^{-22}\,(B/{\rm Gauss})^2$, apparently too
small for any realistic astrophysical $B$-field configuration. For
axion-like particles (ALPs), in contrast, where $m_a$ and
$g_{a\gamma}$ are independent parameters, large conversions and
astrophysical signatures are
conceivable~\cite{Jaeckel:2010ni,Redondo:2010dp}.

The situation does not improve for $a\to\gamma$ conversions near
compact objects such as pulsars or active galactic nuclei where $B$
fields can be much larger and the photon plasma mass can be such that
$q=0$ \cite{Hochmuth:2007hk}. The Hillas diagram of possible sources
of high-energy cosmic rays shows that $B\times L\sim 10 $ G$\times$pc
can be attained. The maximum conversion probability (taking $q\to 0$)
is $(g_{a\gamma} BL/2)^2\sim {\cal O}(1)$.  However, the intrinsic
$\gamma$-ray emission tends to be far too large to disentangle the two
components, even if spectral features could
help~\cite{Chelouche:2008ta}.

\section{Next galactic SN}

The next galactic SN will provide a
high-statistics signal of 10~MeV range neutrinos
\cite{Scholberg:2010zz}. What about the comparable energy release in
100~MeV range axions? The $\bar\nu_ep\to n e^+$ cross section is
$\sigma_{\bar\nu_ep}\sim 9.4\times10^{-44}~{\rm
cm}^2\,(E_{\bar\nu_e}/{\rm MeV})^2$, so for detection energies of
20--30~MeV it is around $10^{-40}~{\rm cm}^2$. For axions, a
reaction like $a+p\to N+\pi$ has a cross section on the $\Delta$
resonance ($E_a\sim 340$~MeV) of order $100~{\rm mb}\,(f_\pi/f_a)^2
\sim 10^{-45}~{\rm cm}^2$. Most of the axion flux is not on
resonance and the rates for such reactions always seem too small for
realistic detection. The largest conceivable SN signal is in a
future megaton detector and if the red supergiant Betelgeuse at a
distance of 200~pc collapses. 
This scenario provides about $3\times10^8$ $\bar\nu_e$
events and conceivably a few events above the tail of the neutrino
spectrum that could be attributed to $ap\to N\pi$. Of course, such
an scenario would require a much more careful discussion.

\section{Next generation axion helioscope}

A more realistic detection
possibility of WD-inspired axions is with a large helioscope beyond
CAST. The conversion probability is $(g_{a\gamma}B L / 2)^2\sim
10^{-20}$ for $L=20$~m, $B=10$~T and $g_{a\gamma}=10^{-12}$
GeV$^{-1}$. The solar axion flux from processes involving electrons
is $0.47\times10^{-6}\,L_\odot$ with an average energy of 2.1~keV
and a flux at Earth of $2.0\times10^9~{\rm cm}^{-2}~{\rm
  s}^{-1}$ \cite{Raffelt:1985nk}, yielding several events per year and
m$^2$. The feasibility of such an instrument with an aperture up to
4~m$^2$ has been recently assessed~\cite{Irastorza:2011gs}.

One amusing application for such an instrument is to detect axions
from a possible Betelgeuse SN explosion. 
Assuming
all SN energy is released in axions of average energy 80~MeV,
Betelgeuse provides an axion fluence at Earth of $5\times
10^{14}$~cm$^{-2}$. 
In this case one needs an aperture
exceeding 20~m$^2$ to get a few events. Pointing the instrument at
Betelgeuse in time for the explosion is possible by the early warning ($\sim$ few days)
provided by the detectable thermal neutrinos from the silicon
burning phase preceding core collapse~\cite{Odrzywolek:2003vn}.

\section{Conclusions}

The intriguing hint from WD cooling for the existence of DFSZ-type
axions with $f_a\sim 10^9$~GeV and $m_a\sim 7$~meV implies that
core-collapse SNe emit a large fraction of their energy as axions.
The universe would be filled with
30~MeV-range axion radiation with a density comparable to the diffuse
SN neutrino background and the extra-galactic background
light.  The axion population produced in the early
universe would comprise only a small fraction of cold dark matter, but
of course the cold dark matter in the universe may well
consist of different components.  Searching for a sub-dominant
meV-mass axion dark matter component is a new challenge that has not
yet been seriously addressed in the literature.  It is intriguing that
axions with such parameters are accessible in a next
generation axion helioscope, a possibility that should be vigorously
pursued. The interpretation of WD cooling in terms of axion emission
is, of course, speculative, but it suggests a fascinating new meV-mass
frontier of axion physics.

\acknowledgments

G.R.\ and J.R.\ acknowledge partial
support by the Deutsche Forschungsgemeinschaft, Grants No.\ TR~27
and EXC~153 (Germany), N.V.M.\ by MECESUP Project PUC0609 (Chile).


\end{document}